\documentclass[11pt]{article}

\usepackage[latin1]{inputenc}\usepackage{epsfig}\usepackage{amsfonts}

\title{
  On the Sums of Inverse Even Powers of \\
  Zeros of Regular Bessel Functions}

\author{
  \Large Jorge L. deLyra \\ Department of Mathematical Physics \\
  Physics Institute \\ University of São Paulo}

\date{February 18, 2013}

\newcommand{\ii}{\imath\,}
\newcommand{\Frac}[2]{\frac{\displaystyle #1}{\displaystyle #2}}

\newcommand{\RR}{{\cal R}}

\begin{document}\maketitle

\begin{abstract}
  \noindent
  We provide a new, simple general proof of the formulas giving the
  infinite sums $\sigma(p,\nu)$ of the inverse even powers $2p$ of the
  zeros $\xi_{\nu k}$ of the regular Bessel functions $J_{\nu}(\xi)$, as
  functions of $\nu$. We also give and prove a general formula for certain
  linear combinations of these sums, which can be used to derive the
  formulas for $\sigma(p,\nu)$ by purely linear-algebraic means, in
  principle for arbitrarily large powers. We prove that these sums are
  always given by a ratio of two polynomials on $\nu$, with integer
  coefficients. We complete the set of known formulas for the smaller
  values of $p$, extend it to $p=9$, and point out a connection with the
  Riemann zeta function, which allows us to calculate some of its values.
\end{abstract}

\section{Introduction}

In boundary value problems involving the diffusion equation the following
infinite sums sometimes appear,

\begin{equation}
  \sigma(p,\nu)
  =
  \sum_{k=1}^{\infty}
  \frac{1}{\xi_{\nu k}^{2p}},
\end{equation}

\noindent
most often for $p=1$, where $\xi_{\nu k}$ are the positions of the zeros
located away from the origin of the regular cylindrical Bessel function
$J_{\nu}(\xi)$, with real $\nu\geq 0$ and integer $p>0$. The sums are
convergent for $p\geq 1$. As we will show in what follows, all these sums
have the property that they are given by the ratio of two polynomials on
$\nu$ with integer coefficients. The simplest and most common example is

\begin{equation}
  \sigma(1,\nu)
  =
  \frac{1}{4(\nu+1)}.
\end{equation}

\noindent
In a few cases the exact expression of these polynomials are available in
the literature~\cite{gnwatson}. The known cases are those obtained by
Rayleigh, extending investigations by Euler, for $p=1$ through $p=5$, and
one discovered by Cayley, for $p=8$. The cases $p=6$ and $p=7$ seem not to
be generally known, and will be given explicitly further on. The known
cases were obtained in a case-by-case fashion, using the expression of the
Bessel functions as infinite products involving its zeros.

In this paper we will provide a simple, independent proof of all the known
formulas, and will present a general formula from which the specific
formulas can be derived, for any given strictly positive integer value of
$p$, by purely algebraic means. The proof will rely entirely on the
general properties of analytical functions and on the well-known
properties of the functions $J_{\nu}(\xi)$, which are generally available
in the literature, for example in~\cite{gradshteyn}.

\section{Definition of the Elements Involved}

The proof will be based on the singularity structure of the following
analytical function in the complex-$\xi$ plane,

\begin{equation}\label{eqnforf}
  f(p,\nu,\xi)
  =
  \frac{J_{\nu+p}(\xi)}{\xi^{p+1}J_{\nu}(\xi)},
\end{equation}

\noindent
where for the time being we may consider that $\nu\geq 0$ and $p>0$ are
real numbers. In case any of the functions involved have branching points
at $\xi=0$, we consider the cuts to be over the negative real semi-axis.
Preliminary to the proof, it will be necessary to establish a few
properties of this function.

\begin{figure}[htp]
  \centering
  \fbox{
    \epsfig{file=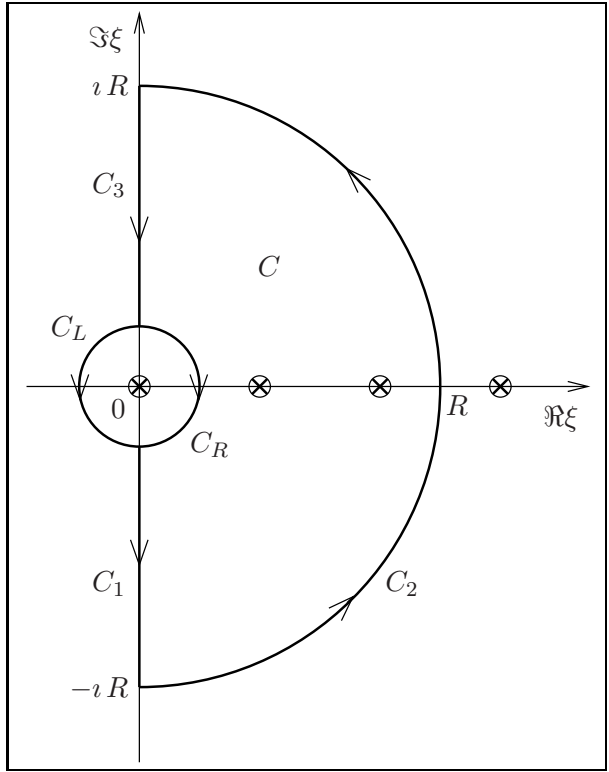,scale=1.0,angle=0}
  }
  \caption{The integration contour in the complex-$\xi$ plane, showing the
    various parts of the circuit $C$ and some of the singularities of
    $f(p,\nu,\xi)$.}
  \label{a06f01}
\end{figure}

We will consider the contour integral of this function over the circuit on
the complex-$\xi$ plane shown in Fig.~\ref{a06f01}, in the $R\to\infty$
limit. Since this circuit goes through the origin $\xi=0$, where the
function will be seen to have a simple pole, we will adopt for the
integral the principal value of Cauchy. The limit $R\to\infty$ will be
taken in a discrete way, in order to avoid going through the other
singularities of the function, which are located at $\xi=\xi_{\nu k}$. We
will see that, for large values of $R$ and $j$, it is possible to adopt
for $R$ the values given by

\begin{equation}\label{eqnforR}
  R
  =
  \pi\,\frac{2\nu+1}{4}
  +
  j\pi,
\end{equation}

\noindent
where for each $k$ there is a value of the integer $j$ such that $R$ is
strictly within the interval $(\xi_{\nu k},\xi_{\nu (k+1)})$. In this way
each step in the discrete $R\to\infty$ limit will correspond to a partial
sum of the infinite sums involved.

The proof of the expressions for $\sigma(p,\nu)$ consists of two parts:
first, the proof that the integral of $f(p,\nu,\xi)$ over the circuit is
zero in the $R\to\infty$ limit, for all $p>0$ and all $\nu\geq 0$; second,
the use of the residue theorem. This will result in a general formula from
which the expressions for the sums $\sigma(p,\nu)$ can be derived. We will
also present a partial solution of the problem of deriving the formulas
for $\sigma(p,\nu)$, which will take the form of another general formula
from which these expressions can be derived algebraically.

\section{Properties of $f(p,\nu,\xi)$}

Let us establish a few important properties of $f(p,\nu,\xi)$, starting by
its behavior under the inversion of the sign of $\xi$. We start with the
analogous property of $J_{\nu}(\xi)$, for which we have, using the
Maclaurin series for these functions~\cite{grads959}, which converges over
the whole complex plane,

\begin{equation}
  J_{\nu}(-\xi)
  =
  (-1)^{\nu}
  J_{\nu}(\xi).
\end{equation}

\noindent
Using this in the expression for $f(p,\nu,\xi)$ we get

\begin{equation}
  f(p,\nu,-\xi)
  =
  -f(p,\nu,\xi),
\end{equation}

\noindent
that is, $f(p,\nu,\xi)$ is an odd function of $\xi$, for all $p$ and all
$\nu$.

Next we show that $f(p,\nu,\xi)$ has a simple pole at $\xi=0$. Since
$J_{\nu+p}(\xi)$, $\xi^{p+1}$ and $J_{\nu}(\xi)$ are analytical functions
over the whole complex-$\xi$ plane, it follows that $f(p,\nu,\xi)$ is
analytical over the whole plane except for those points where the
denominator vanishes, where it has poles. These are the origin $\xi=0$ and
the zeros $\xi_{\nu k}$ of the Bessel function in the denominator. Note
that while for non-integer $\nu$ and $p$ the functions involved have
branching points at $\xi=0$, the function $f(p,\nu,\xi)$ never does. In
order to determine the residue of $f(p,\nu,\xi)$ at $\xi=0$ we consider
the limit

\noindent
\begin{eqnarray}
  r_{0}(p,\nu)
  & = &
  \lim_{\xi\to 0}
  \xi f(p,\nu,\xi)
  \nonumber\\
  & = &
  \frac{\Gamma(\nu+1)}{2^{p}\Gamma(\nu+p+1)},
\end{eqnarray}

\noindent
where $\Gamma(z)$ is the gamma function and we used once more the
Maclaurin series for $J_{\nu}(\xi)$. Since the limit is finite and
non-zero, it follows that $f(p,\nu,\xi)$ has a simple pole at $\xi=0$, and
that $r_{0}(p,\nu)$ is the corresponding residue. Turning to the poles at
$\xi_{\nu k}$, since $\xi_{\nu k}$ is a simple zero of $J_{\nu}(\xi)$, at
which its derivative $J'_{\nu}(\xi)$ is different from zero, it follows
that $f(p,\nu,\xi)$ has a Taylor expansion around this point, with the
form

\begin{equation}
  J_{\nu}(\xi)
  =
  (\xi-\xi_{\nu k})J'_{\nu}(\xi_{\nu k})
  +
  \sum_{i=2}^{\infty}
  c_{i}(\xi-\xi_{\nu k})^{i},
\end{equation}

\noindent
for certain finite coefficients $c_{i}$. In order to determine the residue
of $f(p,\nu,\xi)$ at $\xi_{\nu k}$ we consider then the limit

\noindent
\begin{eqnarray}
  r_{k}(p,\nu)
  & = &
  \lim_{\xi\to\xi_{\nu k}}
  (\xi-\xi_{\nu k})f(p,\nu,\xi)
  \nonumber\\
  & = &
  \frac{J_{\nu+p}(\xi_{\nu k})}{\xi_{\nu k}^{p+1}J'_{\nu}(\xi_{\nu k})},
\end{eqnarray}

\noindent
where we used this Taylor expansion. Since the derivative is finite and
non-zero at $\xi_{\nu k}$, this limit also is finite and non-zero, and
hence it follows that $f(p,\nu,\xi)$ has a simple pole at $\xi_{\nu k}$,
and that $r_{k}(p,\nu)$ is the corresponding residue. We can simplify this
expression using the well-known identity~\cite{grads968}

\begin{equation}
  \xi\frac{\partial}{\partial\xi}J_{\nu}(\xi)
  =
  -\xi J_{\nu+1}(\xi)+\nu J_{\nu}(\xi),
\end{equation}

\noindent
which applied at $\xi=\xi_{\nu k}$, since $J_{\nu}(\xi_{\nu k})=0$,
results in

\begin{equation}
  J'_{\nu}(\xi_{\nu k})
  =
  -J_{\nu+1}(\xi_{\nu k}).
\end{equation}

\noindent
It follows therefore that we have for the residues of the poles at
$\xi_{\nu k}$,

\begin{equation}
  r_{k}(p,\nu)
  =
  -\,\frac{J_{\nu+p}(\xi_{\nu k})}{\xi_{\nu k}^{p+1}J_{\nu+1}(\xi_{\nu k})}.
\end{equation}

We will now establish the behavior of the absolute value of the ratio of
Bessel functions which is contained in the expression of $f(p,\nu,\xi)$ in
Eq.~(\ref{eqnforf}), for large values of $R$. In order to do this we use
the asymptotic expansion of the Bessel functions~\cite{grads961}, valid in
the whole complex plane so long as $\arg(\xi)\neq\pm\pi$, written in terms
of $R=|\xi|$ and $\theta=\arg(\xi)$, to the lowest orders, and with the
trigonometric functions expressed as complex exponentials,

\noindent
\begin{eqnarray}
  J_{\nu}(\xi)
  & = &
  \sqrt{\frac{\,e^{-\ii\theta}}{2\pi R}}
  \left\{
    \left[
      1
      +
      \frac{\RR_{c}(\nu,\xi)}{R^{2}\,e^{\ii 2\theta}}
    \right]
  \right.
  \times
  \nonumber\\
  &   &
  \hspace{4.5em}
  \times
  \left.
    \left[
      e^{-R\sin(\theta)}\,e^{\ii\alpha(R,\theta,\nu)}
      +
      e^{R\sin(\theta)}\,e^{-\ii\alpha(R,\theta,\nu)}
    \right]
  \right.
  +
  \nonumber\\
  &   &
  \hspace{2.7em}
  +
  \ii
  \left.
    \left[
      \frac{4\nu^{2}-1}{8R\,e^{\ii\theta}}
      +
      \frac{\RR_{s}(\nu,\xi)}{R^{3}\,e^{\ii 3\theta}}
    \right]
  \right.
  \times
  \nonumber\\\label{eqnasymp}
  &   &
  \hspace{4.5em}
  \times
  \left.
    \left[
      e^{-R\sin(\theta)}\,e^{\ii\alpha(R,\theta,\nu)}
      -
      e^{R\sin(\theta)}\,e^{-\ii\alpha(R,\theta,\nu)}
    \right]
    \rule{0em}{4ex}
  \right\},
\end{eqnarray}

\noindent
where $\RR_{c}(\nu,\xi)$ and $\RR_{s}(\nu,\xi)$ are certain limited
functions of $\xi$ and $\alpha(R,\theta,\nu)$ is a certain real number,
given by

\begin{equation}
  \alpha(R,\theta,\nu)
  =
  R\cos(\theta)-\pi\,\frac{2\nu+1}{4}.
\end{equation}

\noindent
The behavior of the expression in Eq.~(\ref{eqnasymp}) for large values of
$R$ depends on the sign of $\theta$, and the particular case $\theta=0$
has to be examined separately. In this particular case we have

\noindent
\begin{eqnarray}
  J_{\nu}(\xi)
  & = &
  \sqrt{\frac{2}{\pi R}}
  \left\{
    \left[
      1
      +
      \frac{\RR_{c}(\nu,\xi)}{R^{2}}
    \right]
    \cos[\alpha(R,0,\nu)]
  \right.
  +
  \nonumber\\
  &   &
  \hspace{3.95em}
  -
  \left.
    \left[
      \frac{4\nu^{2}-1}{8R}
      +
      \frac{\RR_{s}(\nu,\xi)}{R^{3}}
    \right]
    \sin[\alpha(R,0,\nu)]
  \right\},
\end{eqnarray}

\noindent
where all the functions involved are now limited, so that for large values
of $R$ we have for the dominant part of $J_{\nu}(\xi)$,

\begin{equation}
  J_{\nu}(\xi)
  \approx
  \sqrt{\frac{2}{\pi R}}\,
  \cos[\alpha(R,0,\nu)].
\end{equation}

\noindent
Note now that the points where $\cos[\alpha(R,0,\nu)]=0$ are the zeros of
$J_{\nu}(\xi)$, expressed in the asymptotic limit. We will now choose a
way to take the $R\to\infty$ limit such that these zeros are avoided. We
may simply chose for the passage of the circuit across the real axis that
point between two zeros where $\cos[\alpha(R,0,\nu)]=\pm 1$ and
$\sin[\alpha(R,0,\nu)]=0$. Since for $\theta=0$ we have

\begin{equation}
  \alpha(R,0,\nu)
  =
  R-\pi\,\frac{2\nu+1}{4},
\end{equation}

\noindent
and we must have $\alpha(R,0,\nu)=j\pi$ for some integer $j$, we conclude
that Eq.~(\ref{eqnforR}) holds, which will cause the crossing of the
circuit and the real axis to avoid the zeros. This defines the
$R\to\infty$ limit in full detail. It follows that for our purposes here
we may write the asymptotic expansion in the case $\theta=0$ as

\begin{equation}
  J_{\nu}(\xi)
  =
  \pm
  \sqrt{\frac{2}{\pi R}}
  \left[
    1
    +
    \frac{\RR_{c}(\nu,\xi)}{R^{2}}
  \right].
\end{equation}

\noindent
In the case $\theta>0$ we put the dominant real exponential in evidence
and obtain

\noindent
\begin{eqnarray}
  J_{\nu}(\xi)
  & = &
  \sqrt{\frac{\,e^{-\ii\theta}}{2\pi R}}\,e^{R\sin(\theta)}
  \left\{
    \left[
      1
      +
      \frac{\RR_{c}(\nu,\xi)}{R^{2}\,e^{\ii 2\theta}}
    \right]
  \right.
  \times
  \nonumber\\
  &   &
  \hspace{8em}
  \times
  \left.
    \left[
      e^{-2R\sin(\theta)}\,e^{\ii\alpha(R,\theta,\nu)}
      +
      e^{-\ii\alpha(R,\theta,\nu)}
    \right]
  \right.
  +
  \nonumber\\
  &   &
  \hspace{6.05em}
  +
  \ii
  \left.
    \left[
      \frac{4\nu^{2}-1}{8R\,e^{\ii\theta}}
      +
      \frac{\RR_{s}(\nu,\xi)}{R^{3}\,e^{\ii 3\theta}}
    \right]
  \right.
  \times
  \nonumber\\
  &   &
  \hspace{8em}
  \times
  \left.
    \left[
      e^{-2R\sin(\theta)}\,e^{\ii\alpha(R,\theta,\nu)}
      -
      e^{-\ii\alpha(R,\theta,\nu)}
    \right]
    \rule{0em}{4ex}
  \right\},
\end{eqnarray}

\noindent
where all the functions within the brackets are now limited or go to zero
in the $R\to\infty$ limit. Finally, we do the same thing for the case
$\theta<0$, obtaining

\noindent
\begin{eqnarray}
  J_{\nu}(\xi)
  & = &
  \sqrt{\frac{\,e^{-\ii\theta}}{2\pi R}}\,e^{-R\sin(\theta)}
  \left\{
    \left[
      1
      +
      \frac{\RR_{c}(\nu,\xi)}{R^{2}\,e^{\ii 2\theta}}
    \right]
  \right.
  \times
  \nonumber\\
  &   &
  \hspace{8.5em}
  \times
  \left.
    \left[
      e^{\ii\alpha(R,\theta,\nu)}
      +
      e^{2R\sin(\theta)}\,e^{-\ii\alpha(R,\theta,\nu)}
    \right]
  \right.
  +
  \nonumber\\
  &   &
  \hspace{6.6em}
  +
  \ii
  \left.
    \left[
      \frac{4\nu^{2}-1}{8R\,e^{\ii\theta}}
      +
      \frac{\RR_{s}(\nu,\xi)}{R^{3}\,e^{\ii 3\theta}}
    \right]
  \right.
  \times
  \nonumber\\
  &   &
  \hspace{8.5em}
  \times
  \left.
    \left[
      e^{\ii\alpha(R,\theta,\nu)}
      -
      e^{2R\sin(\theta)}\,e^{-\ii\alpha(R,\theta,\nu)}
    \right]
    \rule{0em}{4ex}
  \right\},
\end{eqnarray}

\noindent
where once more all the functions within the brackets are now limited or
go to zero in the $R\to\infty$ limit. We are now in a position to analyze
the behavior of the absolute value of the ratio of two Bessel functions
which appears in the definition of $f(p,\nu,\xi)$. The factors which do
not depend on $\nu$ are common to the numerator and denominator, and
cancel out. In the case $\theta=0$ we get

\begin{equation}
  \left|\frac{J_{\nu+p}(\xi)}{J_{\nu}(\xi)}\right|
  =
  \left|
    \frac
    {
      1
      +
      \Frac{\RR_{c}(\nu+p,\xi)}{R^{2}}
    }
    {
      1
      +
      \Frac{\RR_{c}(\nu,\xi)}{R^{2}}
    }
  \right|,
\end{equation}

\noindent
so that in the $R\to\infty$ limit we get

\begin{equation}
  \lim_{R\to\infty}
  \left|\frac{J_{\nu+p}(\xi)}{J_{\nu}(\xi)}\right|
  =
  1.
\end{equation}

\noindent
It is not difficult to verify that for both the case $\theta>0$ and the
case $\theta<0$ we get this same value for this limit. We see therefore
that the $R\to\infty$ limit of the absolute value of this ratio is simply
$1$, for all values of $\theta$ in $(-\pi,\pi)$.

\section{Evaluation of the Integral}

Let us consider now the proof that the integral is zero. In order to do
this we will separate the circuit in sections and prove the result for
each section. The complete circuit $C$ consists of two straight sections
$C_{1}$ e $C_{3}$, of the great semicircle $C_{2}$ and of two small
semicircles $C_{L}$ e $C_{R}$ of radius $\varepsilon$ around the point
$\xi=0$.

For the pair of straight lines $C_{1}$ e $C_{3}$, where we have $d\xi=\ii
dy$ with $\xi=x+\ii y$, taking into account the orientation, we may write

\noindent
\begin{eqnarray}
  I_{C_{1}+C_{3}}
  & = &
  \int_{C_{1}+C_{3}}f(p,\nu,\xi)\,d\xi
  \nonumber\\
  & = &
  \ii
  \int_{R}^{\varepsilon}f(p,\nu,\xi)\,dy
  +
  \ii
  \int_{-\varepsilon}^{-R}f(p,\nu,\xi)\,dy.
\end{eqnarray}

\noindent
Making in the second integral the transformation of variables
$\xi\to-\xi$, which implies $x\to-x$ e $y\to-y$, and since $f(p,\nu,\xi)$
is odd, we have

\begin{equation}
  I_{C_{1}+C_{3}}
  =
  0.
\end{equation}

\noindent
We see therefore that this part of the integral vanishes exactly,
independently of the values of $R$ and $\varepsilon$. We are therefore
free to take limits involving $R$ or $\varepsilon$ during the calculation
of the other sections of the integral, without affecting this result.

Next we consider the two semicircles of radius $\varepsilon$. We will
denote this part of the integral, to be calculated according to the
criterion of the principal value of Cauchy, as

\begin{equation}
  I_{C_{L}+C_{R}}
  =
  \int_{C_{L}+C_{R}}f(p,\nu,\xi)\,d\xi.
\end{equation}

\noindent
Since the function $f(p,\nu,\xi)$ has a simple pole at $\xi=0$, and is
also odd, it can be expressed as a Laurent series around this point, with
the form

\begin{equation}
  f(p,\nu,\xi)
  =
  \frac{r_{0}(p,\nu)}{\xi}
  +
  \sum_{i=0}^{\infty}
  c_{i}\xi^{2i+1},
\end{equation}

\noindent
where $r_{0}(p,\nu)$ is the residue of the function at this point, and
$c_{i}$ are certain finite coefficients. The series is convergent so long
as $\varepsilon$ is smaller than the first zero $\xi_{\nu 1}$. The sum of
positive powers represents an analytical function around $\xi=0$, and is
therefore regular within the circle of radius $\varepsilon$. It follows
that the integral of this regular part goes to zero in the limit
$\varepsilon\to 0$, since in this limit both each individual term of the
sum and the measure of the domain of integration vanish.

It follows that only the integral of the term containing the pole can
remain different from zero in the $\varepsilon\to 0$ limit. We will
therefore calculate this integral in polar coordinates, with
$\xi=\varepsilon\exp(\ii\theta)$ and
$d\xi=\ii\varepsilon\exp(\ii\theta)d\theta$. Since we must use here the
Cauchy principal value we have for this part $\Delta I_{C_{L}+C_{R}}$ of
the integral $I_{C_{L}+C_{R}}$,

\noindent
\begin{eqnarray}
  \Delta I_{C_{L}+C_{R}}
  & = &
  \frac{1}{2}
  \int_{C_{L}}\frac{r_{0}(p,\nu)}{\xi}\,d\xi
  +
  \frac{1}{2}
  \int_{C_{R}}\frac{r_{0}(p,\nu)}{\xi}\,d\xi
  \nonumber\\
  & = &
  \frac{\ii r_{0}(p,\nu)}{2}
  \int_{\pi/2}^{3\pi/2}\,d\theta
  +
  \frac{\ii r_{0}(p,\nu)}{2}
  \int_{\pi/2}^{-\pi/2}\,d\theta
  \nonumber\\
  & = &
  0.
\end{eqnarray}

\noindent
Therefore, this part of the integral $I_{C_{L}+C_{R}}$ also vanishes, and
hence the integral $I_{C_{L}+C_{R}}$ vanishes in the limit $\varepsilon\to
0$. Since during this deformation of the circuit no singularities of the
function are crossed, and hence the integral does not change, if follows
that the integral is zero for all values of $\varepsilon$ smaller than
$\xi_{\nu 1}$.

The last section of the circuit we must consider is $C_{2}$. In this case
the integral is not zero for finite values of $R$, but we may show that it
goes to zero in the limit $R\to\infty$, subject to the condition that for
large values of $R$ we have that Eq.~(\ref{eqnforR}) holds, so that the
circuit does not go over any of the singularities at the points $\xi_{\nu
  k}$. Using once more polar coordinates, in this section of the circuit
we have $d\xi=\ii R\exp(\ii\theta)d\theta$, where $\xi=R\exp(\ii\theta)$,
so that the integral is given by

\noindent
\begin{eqnarray}
  I_{C_{2}}
  & = &
  \int_{C_{2}}f(p,\nu,\xi)\,d\xi
  \nonumber\\
  & = &
  \ii R\int_{-\pi/2}^{\pi/2}d\theta\,e^{\ii\theta}\,f(p,\nu,\xi).
\end{eqnarray}

\noindent
Taking the absolute value of the integral and using the triangle
inequalities we have

\begin{equation}
  |I_{C_{2}}|
  \leq
  R\int_{-\pi/2}^{\pi/2}d\theta\,|f(p,\nu,\xi)|,
\end{equation}

\noindent
for any value of $R$, and hence also in the $R\to\infty$ limit. We must
now consider the behavior of the absolute value of $f(p,\nu,\xi)$ for
large values of $R$. In order to do this we calculate the limit

\noindent
\begin{eqnarray}
  |I_{C_{2}}|
  & \leq &
  \lim_{R\to\infty}
  \left[
    R\int_{-\pi/2}^{\pi/2}d\theta\,
    |f(p,\nu,\xi)|
  \right]
  \nonumber\\
  & = &
  \lim_{R\to\infty}
  \left[
    R\int_{-\pi/2}^{\pi/2}d\theta\,
    \frac{1}{R^{p+1}}\,
    \left|\frac{J_{\nu+p}(\xi)}{J_{\nu}(\xi)}\right|
  \right].
\end{eqnarray}

\noindent
As we established before, the limit of the absolute value of the ratio of
the two Bessel functions is $1$. As a consequence of this, we have for the
integral over the section $C_{2}$ of the circuit, in the $R\to\infty$
limit,

\noindent
\begin{eqnarray}
  |I_{C_{2}}|
  & \leq &
  \lim_{R\to\infty}
  \left[
    \int_{-\pi/2}^{\pi/2}d\theta\,
    \frac{1}{R^{p}}
  \right]
  \nonumber\\
  & = &
  0,
\end{eqnarray}

\noindent
since we have $p>0$. This implies, of course, that $I_{C_{2}}=0$ in the
$R\to\infty$ limit. We see therefore that the integral of $f(p,\nu,\xi)$
over the circuit $C$, in the $R\to\infty$ limit, vanishes in all sections
of the circuit, and hence that the integral is zero in the $R\to\infty$
limit,

\begin{equation}
  \lim_{R\to\infty}
  \oint_{C}f(p,\nu,\xi)\,d\xi
  =
  0.
\end{equation}

\section{Using the Residue Theorem}

Considering that in the $R\to\infty$ limit the poles with residues
$r_{k}(p,\nu)$ are all that exist strictly within the circuit, that the
pole with residue $r_{0}(p,\nu)$ is the only one located over the circuit,
and that the integral is defined as the Cauchy principal value at this
pole, we can use the residue theorem to write for the integral

\begin{equation}
  \lim_{R\to\infty}
  \oint_{C}f(p,\nu,\xi)\,d\xi
  =
  2\pi\ii
  \left[
    \frac{1}{2}\,
    r_{0}(p,\nu)
    +
    \sum_{k=1}^{\infty}
    r_{k}(p,\nu)
  \right].
\end{equation}

\noindent
On the other hand, as we saw above the integral vanishes in the
$R\to\infty$ limit, and hence we have

\begin{equation}
  \frac{1}{2}\,
  r_{0}(p,\nu)
  +
  \sum_{k=1}^{\infty}
  r_{k}(p,\nu)
  =
  0.
\end{equation}

\noindent
We have therefore the following general result involving all these
residues, substituting the values we calculated before for each one of
them,

\begin{equation}\label{eqnResidues}
  \frac{\Gamma(\nu+1)}{2^{p+1}\Gamma(\nu+p+1)}
  =
  \sum_{k=1}^{\infty}
  \frac{1}{\xi_{\nu k}^{p+1}}\,
  \frac{J_{\nu+p}(\xi_{\nu k})}{J_{\nu+1}(\xi_{\nu k})}.
\end{equation}

\noindent
This is valid for any real value of $\nu\geq 0$ and for any real value of
$p>0$.

\section{Proof of Some Known Formulas}

Up to this point $p$ could be any strictly positive real number. From now
on, however, we have to assume that $p$ is a strictly positive {\em
  integer}. In order to further simplify the expression obtained above, in
general it will be necessary to write $J_{\nu+p}(\xi_{\nu k})$ in terms of
$J_{\nu+1}(\xi_{\nu k})$, which can be done using the recurrence formula
of the Bessel functions~\cite{grads967}, so long as $p$ is an integer. Let
us examine a few of the initial cases. For $p=1$ we have simply

\begin{equation}
  \frac{\Gamma(\nu+1)}{2^{2}\Gamma(\nu+2)}
  =
  \sum_{k=1}^{\infty}
  \frac{1}{\xi_{\nu k}^{2}},
\end{equation}

\noindent
so that the formula for the sum that corresponds to this case is

\noindent
\begin{eqnarray}
  \sigma(1,\nu)
  & = &
  \sum_{k=1}^{\infty}
  \frac{1}{\xi_{\nu k}^{2}}
  \nonumber\\
  & = &
  \frac{1}{2^{2}(\nu+1)},
\end{eqnarray}

\noindent
were we used the properties of the gamma function, thus obtaining a
polynomial on $\nu$ in the denominator. In this way we obtain the first of
the known results, and this formula is therefore proven, being valid for
any non-negative real value of $\nu$. For $p=2$ we have

\begin{equation}
  \frac{\Gamma(\nu+1)}{2^{3}\Gamma(\nu+3)}
  =
  \sum_{k=1}^{\infty}
  \frac{1}{\xi_{\nu k}^{3}}\,
  \frac{J_{\nu+2}(\xi_{\nu k})}{J_{\nu+1}(\xi_{\nu k})}.
\end{equation}

\noindent
In order to simplify this expression we write the recurrence formula as

\begin{equation}
  J_{\nu+2}(\xi)
  =
  \frac{2(\nu+1)}{\xi}\,J_{\nu+1}(\xi)-J_{\nu}(\xi),
\end{equation}

\noindent
where we exchanged $\nu$ for $\nu+1$. Applying this for $\xi=\xi_{\nu k}$
and using once more the fact that $J_{\nu}(\xi_{\nu k})=0$, we get

\begin{equation}
  J_{\nu+2}(\xi_{\nu k})
  =
  \frac{2(\nu+1)}{\xi_{\nu k}}\,J_{\nu+1}(\xi_{\nu k}),
\end{equation}

\noindent
so that in this case we have

\begin{equation}
  \frac{\Gamma(\nu+1)}{2^{3}\Gamma(\nu+3)}
  =
  2(\nu+1)
  \sum_{k=1}^{\infty}
  \frac{1}{\xi_{\nu k}^{4}},
\end{equation}

\noindent
from which it follows that the formula for the sum that corresponds to
this case is

\noindent
\begin{eqnarray}
  \sigma(2,\nu)
  & = &
  \sum_{k=1}^{\infty}
  \frac{1}{\xi_{\nu k}^{4}}
  \nonumber\\
  & = &
  \frac{1}{2^{4}(\nu+1)^{2}(\nu+2)}.
\end{eqnarray}

\noindent
We thus obtain the second known result, which is now proven, and which
also has a polynomial on $\nu$ in the denominator. The proof of the first
two formulas is therefore quite straightforward. In the $p=3$ case,
however, something slightly different happens. In this case we have

\begin{equation}
  \frac{\Gamma(\nu+1)}{2^{4}\Gamma(\nu+4)}
  =
  \sum_{k=1}^{\infty}
  \frac{1}{\xi_{\nu k}^{4}}\,
  \frac{J_{\nu+3}(\xi_{\nu k})}{J_{\nu+1}(\xi_{\nu k})},
\end{equation}

\noindent
and hence we must write one more version of the recurrence formula.
Exchanging $\nu$ for $\nu+2$ in the original formula, and applying at
$\xi=\xi_{\nu k}$, we obtain

\begin{equation}
  J_{\nu+3}(\xi_{\nu k})
  =
  \frac
  {2(\nu+2)}
  {\xi_{\nu k}}\,J_{\nu+2}(\xi_{\nu k})-J_{\nu+1}(\xi_{\nu k}).
\end{equation}

\noindent
Substituting in this the solution found for the previous case, which gives
us $J_{\nu+2}(\xi_{\nu k})$ in terms of $J_{\nu+1}(\xi_{\nu k})$, we get

\begin{equation}
  J_{\nu+3}(\xi_{\nu k})
  =
  \left[
    \frac{2^{2}(\nu+1)(\nu+2)}{\xi_{\nu k}^{2}}
    -
    1
  \right]
  J_{\nu+1}(\xi_{\nu k}).
\end{equation}

\noindent
In this way we get in this case the result

\noindent
\begin{eqnarray}
  \frac{\Gamma(\nu+1)}{2^{4}\Gamma(\nu+4)}
  & = &
  \sum_{k=1}^{\infty}
  \frac{1}{\xi_{\nu k}^{4}}
  \left[
    \frac{2^{2}(\nu+1)(\nu+2)}{\xi_{\nu k}^{2}}
    -
    1
  \right]
  \nonumber\\
  & = &
  2^{2}(\nu+1)(\nu+2)
  \sigma(3,\nu)
  -
  \sigma(2,\nu).
\end{eqnarray}

\noindent
We see that in this case a linear combination of the sums of two different
powers of the zeros $\xi_{\nu k}$ appears. Using the properties of the
gamma function and substituting the value obtained previously for
$\sigma(2,\nu)$, we get for $\sigma(3,\nu)$

\begin{equation}
  \sigma(3,\nu)
  =
  \frac{2(\nu+2)}{2^{6}(\nu+1)^{3}(\nu+2)^{2}(\nu+3)}.
\end{equation}

\noindent
The formula for the sum that corresponds to this case is therefore

\noindent
\begin{eqnarray}
  \sigma(3,\nu)
  & = &
  \sum_{k=1}^{\infty}
  \frac{1}{\xi_{\nu k}^{6}}
  \nonumber\\
  & = &
  \frac{1}{2^{5}(\nu+1)^{3}(\nu+2)(\nu+3)},
\end{eqnarray}

\noindent
where we once more have a polynomial on $\nu$ in the denominator. We thus
obtain the third of the known results, and the formula for $\sigma(3,\nu)$
is proven.

It is clear that we can proceed in this way indefinitely, thus obtaining
the formulas for successive values of $p$. In each case it is necessary to
first use the recurrence formula in order to write $J_{\nu+p}(\xi)$ in
terms of $J_{\nu+1}(\xi)$. In general the result will be a linear
combination of sums of several distinct inverse powers of $\xi_{\nu k}$.
At this point the use of the general formula in Eq.~(\ref{eqnResidues})
will produce an expression for the linear combination of the corresponding
sums $\sigma(p,\nu)$. Finally, it is necessary to solve the resulting
expression for the sum with the largest value of $p$ so far, using for
this end the results obtained previously for the other sums. In this way
all the formulas for the sums $\sigma(p,\nu)$ can be derived successively
by purely algebraic means, resulting every time in the ratio of two
polynomials, with the one in denominator completely factored.

\section{Proof of a General Formula}

It is possible to systematize the resolution process described above to
the point where a general formula for the linear combination of the sums
$\sigma(p,\nu)$ can be written. This is based on a systematization of the
general formula for the ratio of Bessel functions, which is found to be

\begin{equation}\label{eqnRatio}
  \frac{J_{\nu+p}(\xi_{\nu k})}{J_{\nu+1}(\xi_{\nu k})}
  =
  \sum_{q=0}^{q_{M}}
  (-1)^{q}\,
  \frac{[(p-1)-q]!\;\Gamma(\nu+p-q)}{[(p-1)-2q]!\;q!\;\Gamma(\nu+q+1)}\,
  \left(\frac{2}{\xi_{\nu k}}\right)^{(p-1)-2q},
\end{equation}

\noindent
where $q_{M}=(p-1)/2$ for odd $p$ and $q_{M}=(p-2)/2$ for even $p$, and
for which we will provide proof in what follows. The use of the general
formula in Eq.~(\ref{eqnResidues}) then produces a corresponding general
formula for the linear combination of the sums $\sigma(p,\nu)$,

\begin{equation}\label{eqnSums}
  \frac{\Gamma(\nu+1)}{2^{p}\Gamma(\nu+p+1)}
  =
  \sum_{q=0}^{q_{M}}
  (-1)^{q}\,
  2^{p-2q}\,
  \frac{[(p-1)-q]!\;\Gamma(\nu+p-q)}{[(p-1)-2q]!\;q!\;\Gamma(\nu+q+1)}\,
  \sigma(p-q,\nu).
\end{equation}

\noindent
Note that the left-hand side of this equation can be written as the
inverse of a polynomial on $\nu$, with integer coefficients, with the
simple use of the properties of the gamma function. On the other hand, the
coefficients on the right-hand side can all be written as polynomials on
$\nu$, with integer coefficients, since we have for the arguments of the
two gamma functions, in the numerator and in the denominator,

\begin{equation}
  (\nu+p-q)
  =
  (\nu+q+1)
  +
  p-2q-1,
\end{equation}

\noindent
where $p-2q-1$ is an integer whose minimum value is $0$ for odd $p$, and
$1$ for even $p$. It follows that, once the equation is solved for
$\sigma(p,\nu)$, resulting in

\noindent
\begin{eqnarray}
  \frac{\Gamma(\nu+p)}{\Gamma(\nu+1)}\,
  \sigma(p,\nu)
  & = &
  \frac{\Gamma(\nu+1)}{2^{2p}\Gamma(\nu+p+1)}
  +
  \nonumber\\
  &   &
  -
  \sum_{q=1}^{q_{M}}
  \frac{(-1)^{q}}{2^{2q}}\,
  \frac{[(p-1)-q]!\;\Gamma(\nu+p-q)}{[(p-1)-2q]!\;q!\;\Gamma(\nu+q+1)}\,
  \sigma(p-q,\nu),
\end{eqnarray}

\noindent
and assuming that the previous sums all have this same property, the
expression for this sum will have the form of the ratio of two polynomials
on $\nu$, with integer coefficients. Hence, since we saw that this is
valid for the first three sums, by finite induction it is valid for {\em
  all} the sums.

This set of equations, taken for all strictly positive integer values of
$p$, forms an infinite linear system of equations in triangular form, that
can be solved iteratively in order to obtain closed forms for
$\sigma(p,\nu)$ in a purely algebraic way, in principle for arbitrary
integer values of $p$, although for large values of $p$ the algebraic work
involved can be very large. However, it is straight, direct algebraic
work, well suited for a computer-algebra approach.

We will now prove these two general formulas. Since the general formula in
Eq.~(\ref{eqnSums}) follows from the general formula in
Eq.~(\ref{eqnRatio}), it suffices to prove the latter. We can do this by
finite induction. Since the upper limits of the summations involved depend
on the parity of $p$, it is necessary to consider the two cases
separately. The first step is to verify that our general formula
reproduces the correct results for the first three cases, which we have
already derived individually. Applying the general formula in
Eq.~(\ref{eqnRatio}) for $p=1$, in which case we have $q_{M}=0$, we obtain
at once

\begin{equation}
  \frac{J_{\nu+1}(\xi_{\nu k})}{J_{\nu+1}(\xi_{\nu k})}
  =
  1,
\end{equation}

\noindent
which is obviously the correct result. Applying now the same general
formula for $p=2$, for which we also have $q_{M}=0$, we get

\begin{equation}
  \frac{J_{\nu+2}(\xi_{\nu k})}{J_{\nu+1}(\xi_{\nu k})}
  =
  \frac{2(\nu+1)}{\xi_{\nu k}},
\end{equation}

\noindent
which is also the correct result. Finally, applying the general formula
for $p=3$, in which case we have $q_{M}=1$, we obtain

\begin{equation}
  \frac{J_{\nu+3}(\xi_{\nu k})}{J_{\nu+1}(\xi_{\nu k})}
  =
  \frac{2^{2}(\nu+2)(\nu+1)}{\xi_{\nu k}^{2}}
  -
  1,
\end{equation}

\noindent
which once more is the correct result. It suffices now to use the
recurrence formula of the Bessel functions to show that the formula for
$p$ follows from the previous formulas, for $p-1$ e $p-2$. We start with
the case in which $p$ is even, and writing explicitly the upper limits of
the sums, we have

\noindent
\begin{eqnarray}\label{eqnEM1}
  \frac{J_{\nu+p-1}(\xi_{\nu k})}{J_{\nu+1}(\xi_{\nu k})}
  & = &
  \sum_{q=0}^{(p-2)/2}
  (-1)^{q}\,
  \frac{[(p-2)-q]!\;\Gamma(\nu+p-1-q)}{[(p-2)-2q]!\;q!\;\Gamma(\nu+q+1)}\,
  \left(\frac{2}{\xi_{\nu k}}\right)^{(p-2)-2q},
  \\\label{eqnEM2}
  \frac{J_{\nu+p-2}(\xi_{\nu k})}{J_{\nu+1}(\xi_{\nu k})}
  & = &
  \sum_{q=0}^{(p-4)/2}
  (-1)^{q}\,
  \frac{[(p-3)-q]!\;\Gamma(\nu+p-2-q)}{[(p-3)-2q]!\;q!\;\Gamma(\nu+q+1)}\,
  \left(\frac{2}{\xi_{\nu k}}\right)^{(p-3)-2q}.
\end{eqnarray}

\noindent
Writing now the recurrence formula which gives the function
$J_{\nu+p}(\xi_{\nu k})$ in terms of $J_{\nu+p-1}(\xi_{\nu k})$ and
$J_{\nu+p-2}(\xi_{\nu k})$, and substituting Eq.~(\ref{eqnEM1})
and~(\ref{eqnEM2}), we get, after some manipulation of the indices of the
sums,

\noindent
\begin{eqnarray}
  J_{\nu+p}(\xi_{\nu k})
  & = &
  \frac{2(\nu+p-1)}{\xi_{\nu k}}\,J_{\nu+p-1}(\xi_{\nu k})
  -
  J_{\nu+p-2}(\xi_{\nu k})
  \Rightarrow
  \nonumber\\
  \frac{J_{\nu+p}(\xi_{\nu k})}{J_{\nu+1}(\xi_{\nu k})}
  & = &
  \frac{2(\nu+p-1)}{\xi_{\nu k}}\,
  \frac{J_{\nu+p-1}(\xi_{\nu k})}{J_{\nu+1}(\xi_{\nu k})}
  -
  \frac{J_{\nu+p-2}(\xi_{\nu k})}{J_{\nu+1}(\xi_{\nu k})}
  \nonumber\\
  & = &
  \sum_{q=0}^{(p-2)/2}
  (-1)^{q}\,
  \frac
  {[(p-1)-q]!\;\Gamma(\nu+p-q)}
  {[(p-1)-2q]!\;q!\;\Gamma(\nu+q+1)}
  \times
  \nonumber\\
  &   &
  \hspace{3em}
  \times
  \frac
  {(\nu+p-1)[(p-1)-2q]+q(\nu+q)}
  {[(p-1)-q][\nu+(p-1)-q]}
  \left(\frac{2}{\xi_{\nu k}}\right)^{(p-1)-2q}.
\end{eqnarray}

\noindent
It is easy to verify that we have for the second fraction in this sum,

\begin{equation}
  \frac
  {(\nu+p-1)[(p-1)-2q]+q(\nu+q)}
  {[(p-1)-q][\nu+(p-1)-q]}
  =
  1.
\end{equation}

\noindent
We therefore conclude that

\begin{equation}
  \frac{J_{\nu+p}(\xi_{\nu k})}{J_{\nu+1}(\xi_{\nu k})}
  =
  \sum_{q=0}^{(p-2)/2}
  (-1)^{q}\,
  \frac{[(p-1)-q]!\;\Gamma(\nu+p-q)}{[(p-1)-2q]!\;q!\;\Gamma(\nu+q+1)}\,
  \left(\frac{2}{\xi_{\nu k}}\right)^{(p-1)-2q},
\end{equation}

\noindent
thus proving the general formula for even $p$. For odd $p$, once more
writing explicitly the upper limit of the sums, we start from

\noindent
\begin{eqnarray}\label{eqnOM1}
  \frac{J_{\nu+p-1}(\xi_{\nu k})}{J_{\nu+1}(\xi_{\nu k})}
  & = &
  \sum_{q=0}^{(p-3)/2}
  (-1)^{q}\,
  \frac{[(p-2)-q]!\;\Gamma(\nu+p-1-q)}{[(p-2)-2q]!\;q!\;\Gamma(\nu+q+1)}\,
  \left(\frac{2}{\xi_{\nu k}}\right)^{(p-2)-2q},
  \\\label{eqnOM2}
  \frac{J_{\nu+p-2}(\xi_{\nu k})}{J_{\nu+1}(\xi_{\nu k})}
  & = &
  \sum_{q=0}^{(p-3)/2}
  (-1)^{q}\,
  \frac{[(p-3)-q]!\;\Gamma(\nu+p-2-q)}{[(p-3)-2q]!\;q!\;\Gamma(\nu+q+1)}\,
  \left(\frac{2}{\xi_{\nu k}}\right)^{(p-3)-2q}.
\end{eqnarray}

\noindent
Writing once again the recurrence formula which gives $J_{\nu+p}(\xi_{\nu
  k})$ in terms of $J_{\nu+p-1}(\xi_{\nu k})$ and $J_{\nu+p-2}(\xi_{\nu
  k})$, and substituting Eq.~(\ref{eqnOM1}) and~(\ref{eqnOM2}), we get,
after some similar manipulation of the indices of the sums,

\noindent
\begin{eqnarray}
  J_{\nu+p}(\xi_{\nu k})
  & = &
  \frac{2(\nu+p-1)}{\xi_{\nu k}}\,J_{\nu+p-1}(\xi_{\nu k})
  -
  J_{\nu+p-2}(\xi_{\nu k})
  \Rightarrow
  \nonumber\\
  \frac{J_{\nu+p}(\xi_{\nu k})}{J_{\nu+1}(\xi_{\nu k})}
  & = &
  \frac{2(\nu+p-1)}{\xi_{\nu k}}\,
  \frac{J_{\nu+p-1}(\xi_{\nu k})}{J_{\nu+1}(\xi_{\nu k})}
  -
  \frac{J_{\nu+p-2}(\xi_{\nu k})}{J_{\nu+1}(\xi_{\nu k})}
  \nonumber\\
  & = &
  \sum_{q=0}^{(p-3)/2}
  (-1)^{q}\,
  \frac
  {[(p-1)-q]!\;\Gamma(\nu+p-q)}
  {[(p-1)-2q]!\;q!\;\Gamma(\nu+q+1)}
  \times
  \nonumber\\
  &   &
  \hspace{3em}
  \times
  \frac
  {(\nu+p-1)[(p-1)-2q]+q(\nu+q)}
  {[(p-1)-q][\nu+(p-1)-q]}
  \left(\frac{2}{\xi_{\nu k}}\right)^{(p-1)-2q}
  \nonumber\\
  &   &
  +
  (-1)^{(p-1)/2}.
\end{eqnarray}

\noindent
The second fraction within the last summation, which does not involve
factorials, is the same as before, and therefore is equal to $1$. It
follows that we have

\noindent
\begin{eqnarray}
  \frac{J_{\nu+p}(\xi_{\nu k})}{J_{\nu+1}(\xi_{\nu k})}
  & = &
  \sum_{q=0}^{(p-3)/2}
  (-1)^{q}\,
  \frac
  {[(p-1)-q]!\;\Gamma(\nu+p-q)}
  {[(p-1)-2q]!\;q!\;\Gamma(\nu+q+1)}\,
  \left(\frac{2}{\xi_{\nu k}}\right)^{(p-1)-2q}
  \nonumber\\
  &   &
  +
  (-1)^{(p-1)/2}.
\end{eqnarray}

\noindent
It is not difficult to verify that the additional term that we have here
is in fact equal to the argument of the summation in the case $q=(p-1)/2$,
so that we may merge it with the summation and this obtain

\begin{equation}
  \frac{J_{\nu+p}(\xi_{\nu k})}{J_{\nu+1}(\xi_{\nu k})}
  =
  \sum_{q=0}^{(p-1)/2}
  (-1)^{q}\,
  \frac{[(p-1)-q]!\;\Gamma(\nu+p-q)}{[(p-1)-2q]!\;q!\;\Gamma(\nu+q+1)}\,
  \left(\frac{2}{\xi_{\nu k}}\right)^{(p-1)-2q},
\end{equation}

\noindent
thus proving the general formula in this case. This completes the proof of
the general formula in Eq.~(\ref{eqnRatio}), from which follows the
general formula in Eq.~(\ref{eqnSums}) for the linear combination of the
sums $\sigma(p,\nu)$, which is therefore proven as well.

\section{Some Particular Cases}

We may now use the general formula in Eq.~(\ref{eqnSums}) for
$\sigma(p,\nu)$ in order to write explicitly a few cases which are not
found in the current literature. With a little help from the free-software
algebraic manipulation program {\tt maxima}, we get the following two
results, thus completing the sequence of known results up to $p=8$,

\noindent
\begin{eqnarray}
  \sigma(6,\nu)
  & = &
  \frac
  {21\nu^{3}+181\nu^{2}+513\nu+473}
  {2^{11}(\nu+1)^{6}(\nu+2)^{3}(\nu+3)^{2}(\nu+4)(\nu+5)(\nu+6)},
  \\
  \sigma(7,\nu)
  & = &
  \frac
  {33\nu^{3}+329\nu^{2}+1081\nu+1145}
  {2^{12}(\nu+1)^{7}(\nu+2)^{3}(\nu+3)^{2}(\nu+4)(\nu+5)(\nu+6)(\nu+7)}.
\end{eqnarray}

\noindent
We now point out that our results for $\sigma(p,\nu)$ are valid for all
real values of $\nu$, not just for the integers. Therefore, exchanging
$\nu$ for $\nu+1/2$ we may obtain formulas that are valid for the zeros of
the regular spherical Bessel functions $j_{\nu}(\xi)$, since we have the
well-known relation

\begin{equation}
  j_{\nu}(\xi)
  =
  \sqrt{\frac{\pi}{2\xi}}\,
  J_{\nu+1/2}(\xi)
\end{equation}

\noindent
between these two families of functions. In particular, using the value
$\nu=1/2$ we obtain the results for $j_{0}(\xi)$, whose zeros are given by
$k\pi$, since this particular function is proportional to
$\sin(\xi)$~\cite{grads965}. In this way we obtain a direct relation
between our results and the Riemann zeta function, for certain real
integer arguments of $\zeta(z)$. In fact, we have

\noindent
\begin{eqnarray}
  \sigma(p,1/2)
  & = &
  \sum_{k=1}^{\infty}\frac{1}{\pi^{2p}k^ {2p}}
  \Rightarrow
  \nonumber\\
  \zeta(2p)
  & = &
  \pi^{2p}\sigma(p,1/2).
\end{eqnarray}

\noindent
Using the formulas we obtained here for $\sigma(p,\nu)$ in the case
$\nu=1/2$, we obtain for example the values

\noindent
\begin{eqnarray}
  \zeta(12)
  & = &
  \frac{691\pi^{12}}{3^{6}\times 5^{3}\times 7^{2}\times 11\times 13},
  \\
  \zeta(14)
  & = &
  \frac{2\pi^{14}}{3^{6}\times 5^{2}\times 7\times 11\times 13}.
\end{eqnarray}

\noindent
Finally, using our general formula for the case $p=9$ and solving for
$\sigma(9,\nu)$, once more with some help from the free-software program
{\tt maxima}, we obtain

\begin{equation}
  \sigma(9,\nu)
  =
  \frac{P_{9}(\nu)}{Q_{9}(\nu)},
\end{equation}

\noindent
where the two polynomials are given by

\noindent
\begin{eqnarray}
  P_{9}(\nu)
  & = &
  715\,\nu^{6}
  +
  16567\,\nu^{5}
  +
  158568\,\nu^{4}
  +
  798074\,\nu^{3}
  +
  \nonumber\\
  &   &
  +
  2217079\,\nu^{2}
  +
  3212847\,\nu
  +
  1893046,
  \\
  Q_{9}(\nu)
  & = &
  2^{17}
  (\nu+1)^{9}
  (\nu+2)^{4}
  (\nu+3)^{3}
  (\nu+4)^{2}
  \times
  \nonumber\\
  &   &
  \times
  (\nu+5)
  (\nu+6)
  (\nu+7)
  (\nu+8)
  (\nu+9).
\end{eqnarray}

\noindent
Up to the case $p=8$ it is possible and in fact fairly easy to verify the
resulting formulas numerically with good precision, with the use of
standard computational facilities. However, in this $p=9$ case it is just
too difficult to verify this formula by numerical means, except for
$\nu=0$, using the usual double-precision floating-point arithmetic. In
order to do this one would have to use quadruple precision or better
numerical arithmetic. The difficulty seems to lie in the direct numerical
calculation of the sum $\sigma(9,\nu)$, not in the evaluation of the ratio
of polynomials. Hence, the results discussed here acquire an algorithmic,
numerical significance, enabling one to easily calculate the values of the
sums.

\section{Acknowledgements}

The author would like to thank his friend and colleague Prof. Carlos
Eugênio Imbassay Carneiro, for all his interest and help, as well as his
helpful criticism regarding this work.

The author would like to thank Prof. Martin E. Muldoon, of the Department
of Mathematics and Statistics of York University, for drawing his
attention to the fact that there is a whole mathematical literature on
these sums, which are known in that literature as Rayleigh functions.

\section{Historical Note}

From the information supplied by Prof. Muldoon it seems that, since the
time of Euler, aspects of the subject of these sums has been discovered
and rediscovered, possibly more than just once. The explicit expressions
of the sums as functions of $\nu$ were given, for the first $12$ cases, by
Lehmer~\cite{lehmer}. The relation of the sums to the diffusion equation
was first established by Kapitsa~\cite{kapitsa1,kapitsa2}, who also worked
on the calculation of the sums. The recursive general solution presented
here seems to be equivalent to a known recursion formula, first derived by
Meiman~\cite{meiman} and later rediscovered by Kishore~\cite{kishore}.

\end{document}